\documentclass[12pt]{emulateapj}

\pdfoutput=1

\usepackage{graphicx}
\usepackage{amssymb}
\usepackage{natbib}

\slugcomment{Accepted for publication in The Astronomical Journal (2008 August 21)}

\shorttitle{Colors of CKBOs}
\shortauthors{Peixinho, Lacerda and Jewitt}

\begin{document}

\title{Color-Inclination Relation of the Classical Kuiper Belt Objects}

\author{Nuno Peixinho\altaffilmark{1,2,3}, Pedro Lacerda\altaffilmark{1}, and David Jewitt\altaffilmark{1}}
\affil{1 Institute for Astronomy, University of Hawaii, \\
2680 Woodlawn Drive, Honolulu, HI 96822, U.S.A.}
\affil{2 Centre for Computational Physics, University of Coimbra, P-3004-516 Coimbra, Portugal}
\affil{3 Astronomical Observatory of the University of Coimbra, P-3040-004 Coimbra, Portugal}
\email{peixinho@ifa.hawaii.edu}

\begin{abstract}

We re-examine the correlation between the colors and the inclinations of the
Classical Kuiper Belt Objects (CKBOs) with an enlarged sample of optical
measurements. The correlation is strong ($\rho=-0.7$) and highly significant ($>8\,\sigma$) 
in the range $0^{\circ}-34^{\circ}$. Nonetheless, the optical colors are independent of inclination below
$\approx 12^{\circ}$, showing no evidence for a break at the reported boundary
between the so-called dynamically ``hot'' and ``cold'' populations near
$\approx 5^{\circ}$. The commonly accepted parity between the dynamically cold CKBOs
and the red CKBOs is observationally unsubstantiated, since the group of red CKBOs 
extends to higher inclinations. Our data suggest, however, the existence of a different color break. 
We find that the functional form
of the color-inclination relation is most satisfactorily described by a non-linear
and stepwise behavior with a color break at $\approx 12^{\circ}$. 
Objects with inclinations $\geqslant 12^{\circ}$ show bluish
colors which are either weakly correlated with inclination or are simply
homogeneously blue, whereas objects with inclinations $<12^{\circ}$ are homogeneously red. 

\end{abstract}



\section{Introduction}

The Kuiper Belt is a disk of icy bodies having semi-major axes larger than that
of Neptune. Its members are usually known as Kuiper Belt Objects (KBOs) or
Trans-Neptunian Objects (TNOs). The distribution of their orbits is structured
leading to the identification of several dynamical families.  Resonant KBOs are
those which are trapped in mean-motion resonances with Neptune (those trapped
in the 3:2 resonance are also known as Plutinos).  Scattered KBOs, also known
as Scattered Disk Objects, are essentially highly eccentric KBOs under strong
gravitational influence of Neptune. Classical KBOs (CKBOs) possess relatively
circular orbits that are neither located in any strong mean-motion resonance
with Neptune nor strongly subject to its gravitational influence. 

Since their discovery in 1992 more than 1200 KBOs have been identified. Due to
their faintness only about 50 can be spectroscopically studied with the currently
available instruments.  Multicolor photometry provides, however, a first-order
approximation of their spectra, hence of their surface composition.  Most KBOs
can be studied photometrically and about 230 objects have at least one measured
color. Their surface colors have shown to be most diverse, ranging
from neutral and even slightly blue (relative to the Sun) to extremely red,
suggesting a large compositional diversity \citep[see review by][]{Dor+08}. 

The origin of the color diversity remains unclear. Various suggestions have
been made in the context of collisional resurfacing \citep{LuuJew96, GilH02,
Del+04}. Nevertheless, none of the proposed models
has been able to consistently explain the colors \citep{JewLuu01, TheDor03,
Del+04}.  Another possibility is that the observed color differences reflect
primordial compositional variations \citep[e.g.][]{TRC03}. Such
compositional differences would be hard to explain if KBOs formed {\it in
situ}, since the temperature difference between 30 and 50 AU is a very modest
$\approx 10$ K. However, larger temperature and compositional differences might be
possible if the KBO population, or part of it, did not form in place. Some dynamical
models, in fact, suggest outward migration of KBOs \citep[e.g.][]{Malho95,
Gomes03}. Although there are no detailed chemical studies to address how varied
such compositions would be and how these would reflect in the surface colors,
these dynamical models imply a link between the current orbital inclinations of
classical KBOs and their presumed location of origin. 

As a whole, KBOs do not show significant correlations between their colors and
orbital parameters such as semi-major axis or perihelion distance. On the other
hand, the CKBOs do show a correlation between orbital inclination and optical
color \citep{TegRom00, TruBro02} and a correlation between perihelion and
color \citep{TegRom00, Peix+04}.
In parallel, several works have pointed out the existence of two groups of
CKBOs with a separation at $\approx 5^{\circ}$ in inclination. The two groups,
usually referred to as ``cold'' ($i\lesssim 5^{\circ}$) and ``hot'' ($i\gtrsim
5^{\circ}$), have been identified using not only orbital properties
\citep{Brown01, Ell+05} but also physical properties such as size and binarity
\citep{LS01,Peix+04,Gul+06, Noll+08}. Given its potential importance, we
re-examine the color-inclination relation using a new data set and appropriate
statistical tests. 


\section{Data Set}
\label {dataset:sec}

We use the system of \cite{LykMuk07} to select CKBOs for our sample.  These
authors classify the KBO families based on 4 Gyr dynamical simulations.  
The Classical KBOs have semi-major axes in the range $37<a<48$
AU and perihelion distances $q>37$ AU, and must not be not located in
any strong resonance (3:2, 5:3, 7:4, and 2:1). All orbital elements were
gathered from the Minor Planet
Center\footnote{http://cfa-www.harvard.edu/iau/lists/TNOs.html}. We have
computed the orbital inclinations relative to the Kuiper Belt Plane (KBP),
hereafter denoted by $i_k$, as defined by \cite{Ell+05}.   We conducted our
statistical tests using both the raw inclinations and $i_k$, finding no
significant difference between them.  In the remainder of this work, we present
all results in terms of $i_k$.

For our data set we have gathered all the $B-R$ colors of CKBOs available in
the literature or online. Several objects have colors reported in more than one
work but the different measurements have been shown to be essentially
compatible. Hence we chose to take the CKBO colors sequentially from the works
that have been presenting lower and less dispersed error bars to those with
(slightly) larger and more dispersed error bars. Therefore, firstly we gathered
the CKBO colors from Tegler, Romanishin and Consolmagno's data
sample\footnote{http://www.physics.nau.edu/$\sim$tegler/research/survey.htm}
\citep[][and references therein]{TRC03}; secondly those from the ``ESO Large
Program on Centaurs and TNOs" \citep[][and references therein]{Peix+04};
thirdly those from the ``Meudon Multicolor Survey (2MS)" \citep[][and
references therein]{Dor+05}; fourthly those from \cite{JewPeiHsi07}; and lastly
those from the online \cite{HaiDel02} MBOSS
database\footnote{http://www.sc.eso.org/$\sim$ohainaut/MBOSS/}. 
The resulting CKBO sample used here has colors in the range $0.99 \leqslant B-R \leqslant
1.94$.  For reference, the color of the Sun is $(B-R)_{\sun}=0.99$
\citep{Hart+90}.

A histogram of the $B-R$ error bars of the gathered CKBOs shows a rather
continuous but very skewed distribution, from 0.01 up to 0.21 peaking around
0.06 and with a mean value of 0.09. Six wayward objects possess errors between
0.28 and 0.40, though, and we chose to eliminate them. The subsequent data
consist of the $B-R$ colors of 71 CKBOs (see Table \ref{data.tab}).
As discussed in the next section, two data points appear to be
outliers: 2001QY$_{297}$ and 1998WV$_{24}$. 
We chose to discard both and the final data set under analysis
consists of the $B-R$ colors of 69 CKBOs. The effects of keeping these two wayward objects in the sample are discussed in Section \ref{checking:sec}.


\section{Data Analysis}
\label{dataanalysis:sec}

\begin{figure}
\epsscale{1.}
\begin{center}
\plotone{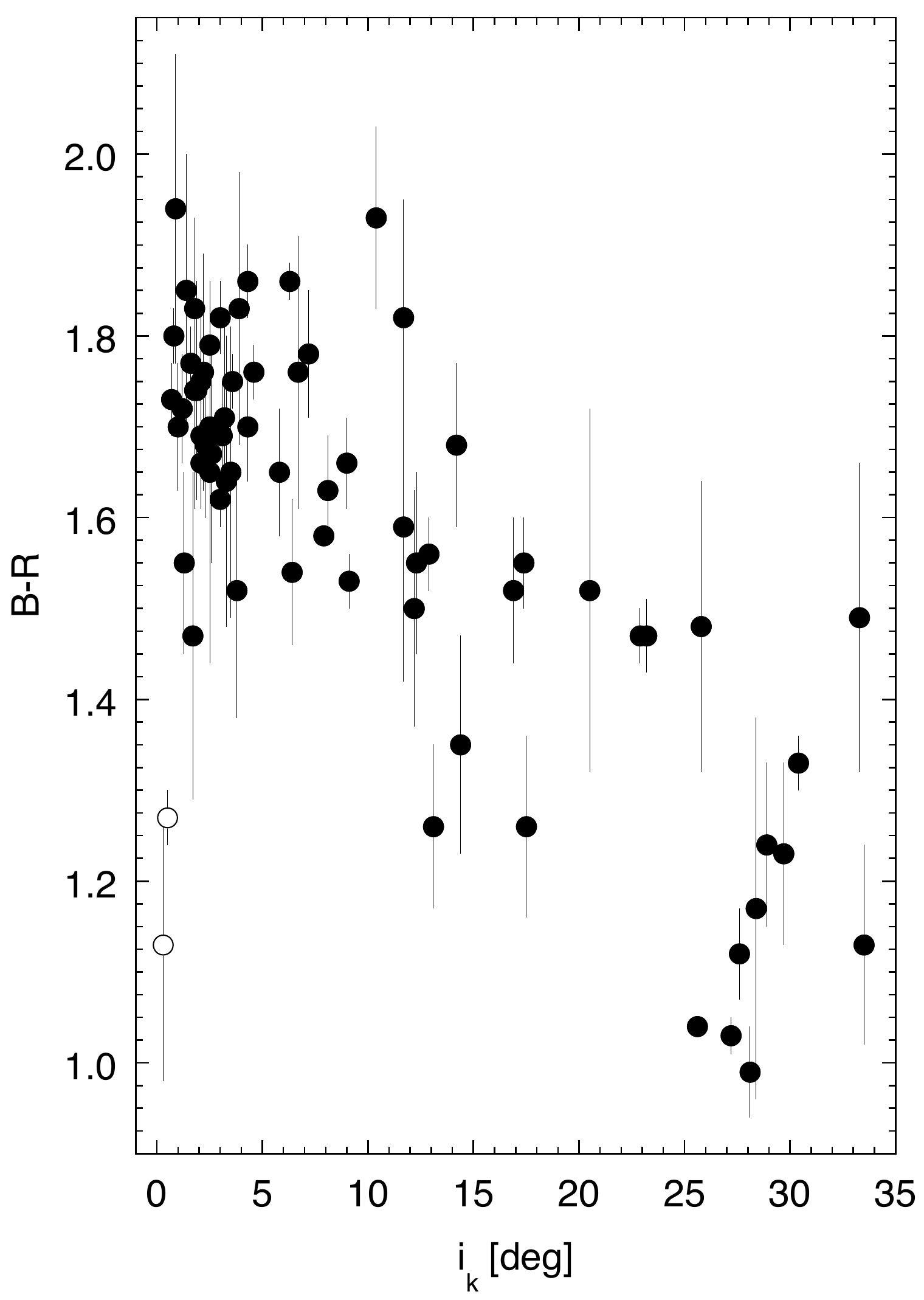}
\caption{$B-R$ colors versus inclination to the Kuiper Belt Plane, $i_k$[deg],
of our data set of 69 CKBOs (black dots). The two apparent outliers that were
discarded from our analysis are indicated (empty circles).\label{plot_ik_br.fig}}
\end{center} 
\end{figure}

A visual inspection of the $B-R$ colors of CKBOs versus $i_k$ in our data set
instantly shows a trend between these two variables (see
Fig.~\ref{plot_ik_br.fig}). Using the statistical tools implemented in IDL we have 
analyzed this trend quantitatively.  
The Spearman-rank correlation coefficient, $\rho$, for the total of N=69 CKBOs
\citep{Spe04} is:

\begin{equation} \label{rhototal:eq}
\rho=-0.70^{+0.09}_{-0.07} \quad SL>8\,\sigma
\end{equation}

\noindent where $SL$ is the significance level in standard deviations of a
Gaussian probability distribution --- error bars are estimated from 1000 bootstrap extractions
corrected for non-Gaussian behavior \citep{EfrTib93}. This is a highly significant correlation
consistent with the published values. The square of the correlation
coefficient, usually called the ``coefficient of determination'', gives
approximately the proportion of the variation in the dependent variable that
can be predicted by the changes in the values of the independent variable. So,
from $\rho^2=(-0.70)^2=0.49$ we may say that about half of the color
variability can be accounted for by differences in orbital inclination. The
other half is color variability unaccounted for by inclination differences and
presumably related to some other undetermined variable or effect. 

Fig.~\ref{plot_ik_br.fig} suggests that the color-inclination trend might be
not linear: the colors of low inclination objects do not seem to correlate with
inclination. When dividing the data set in two groups in inclination with equal
number of objects we have a low inclination group with 34 objects
($i_k<5^{\circ}$) and a high inclination group with 35 ($i_k\geqslant
5^{\circ}$). While the high inclination group still shows a strong and
significant color-inclination correlation: $\rho=-0.81^{+0.05}_{-0.04}$
($SL>8\,\sigma$), the low inclination one does not show any significant
correlation: $\rho=-0.13^{+0.21}_{-0.20}$ ($SL=0.7\,\sigma$). 

\begin{figure}
\epsscale{0.8}
\begin{center}
\plotone{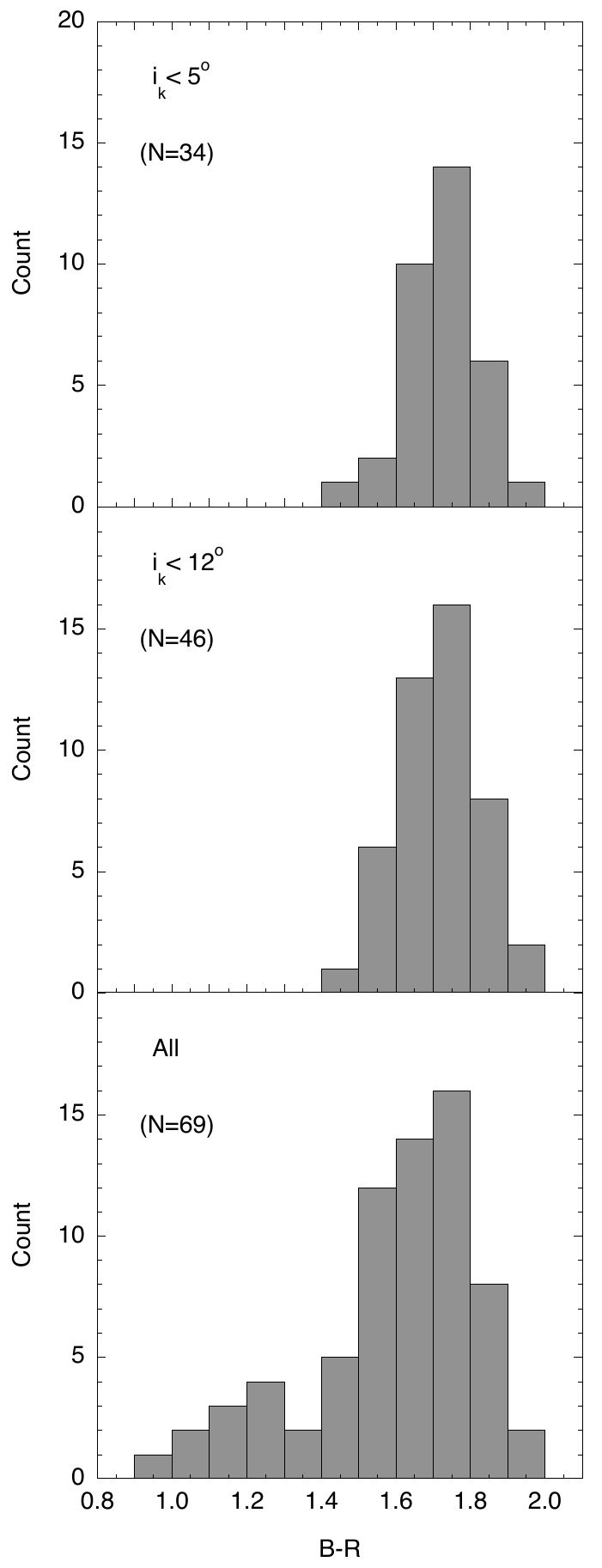}
\caption{Histograms of $B-R$ colors with $i_k<5^{\circ}$ (top), 
with $i_k<12^{\circ}$ (middle), and all $i_k$ values (bottom). 
Only for $i_k\geqslant12^{\circ}$ do we see a
significant number of blue objects, whereas objects between $5^{\circ}$ and
$12^{\circ}$ do not appear different from those with
$i_k<5^{\circ}$.\label{hist_br_ik_stacked.fig}}
\end{center} 
\end{figure}

In Fig.~\ref{hist_br_ik_stacked.fig} we have drawn histograms of the $B-R$
colors for the 34 objects with  $i_k<5^{\circ}$, for the 46 objects with
$i_k<12^{\circ}$, and for all objects. From this figure it seems that the color distributions for
$i_k<5^{\circ}$ and $i_k<12^{\circ}$ are the same while only for $i_k \geqslant
12^{\circ}$ do we start to see a significant number of blue objects. The color 
differences between two groups of objects may be analyzed using the 
Wilcoxon Test \citep{Wilcoxon45}, 
the non-parametric equivalent of the t-Test, also known as Wilcoxon
Rank-Sum Test. The test ranks the full set of colors and assesses for
incompatibility by comparing the ranks assigned to the members of each group. 
Comparing the 34 objects with $i_k<5^{\circ}$ and the 12 objects with 
$5\leqslant i_k<12^{\circ}$ shows no evidence for color differences between the two groups (the significance level of incompatibility is $0.8\,\sigma$). On the other hand, comparing 
the 46 objects with $i_k<12^{\circ}$ and the 23 objects with $i_k \geqslant
12^{\circ}$ shows a color incompatibility at a $6.3\,\sigma$ significance level. 
Next we study: (i) how the correlation coefficient varies with the inclusion of
more highly inclined objects, (ii) how the mean colors vary with inclination,
and (iii) which functional form best describes their behavior.

\subsection{Correlation as a function of inclination}
\label{corrinc:sec}

To further investigate the variations of the color-inclination trend we have
successively computed $\rho$ for CKBOs below a critical inclination cutoff
$i_k^c$ varying from $3^{\circ}$ in increments of $0.5^{\circ}$ up to
$20^{\circ}$. These two extrema were imposed so as not to calculate
correlation values for very small sub-samples which were already out of the
region of interest. Table \ref{rhos_wilcoxon.tab} lists the results for each
inclination cutoff $i_k^c$, both for objects with inclinations below $i_k^c$
and those above $i_k^c$ --- error bars and significance levels are also
indicated. 
We see that $\rho$ varies rather erratically until
$i_{k}^c=12^{\circ}$, increases systematically with the inclusion of
objects above that point, and reaches the $2\,\sigma$ ($95\%$) typical minimum
statistical threshold to have ``reasonably strong evidence'' for correlation at
$i_{k}^c=13.5^{\circ}$.  Such behavior suggests the presence of a homogenous
set of colors below $i_{k}\approx12^{\circ}-13.5^{\circ}$ consistent with 
Fig.~\ref{hist_br_ik_stacked.fig}. While for the 46
CKBOs with $i_{k}< 12^{\circ}$ we have no apparent color-inclination
correlation ($\rho=-0.15^{+0.18}_{-0.17}$, $SL=1.0\,\sigma$), for the 23
objects with $i_{k}\geqslant 12^{\circ}$ a significant correlation is present
($\rho=-0.62^{+0.14}_{-0.11}$, $SL=3.2\,\sigma$). We note that after moving the
critical inclination cutoff by just $0.5^{\circ}$ the 21 CKBOs with
$i_{k}\geqslant 12.5^{\circ}$ no longer show the canonical $3\,\sigma$ level
correlation ($\rho=-0.55^{+0.18}_{-0.14}$, $SL=2.6\,\sigma$), and for the 17
objects with $i_{k}\geqslant 14.5^{\circ}$ the significance level drops below
$2\,\sigma$ ($\rho=-0.45^{+0.28}_{-0.21}$).  Thus, the data provide no formally
significant evidence for correlation among objects with $i_{k}\geqslant
14.5^{\circ}$.

This first analysis shows that the CKBOs of smallest inclination are
homogeneous and red, as other works have reported, but that homogeneity extends 
at least up to $i_{k}\approx12^{\circ}-13.5^{\circ}$, not only up to
$i_{k}\approx5^{\circ}$. Further, the rapid decrease in the correlation found
by removing the objects between $12^{\circ}$ and $14^{\circ}$ may suggest 
two separate groups of objects, each one
having no color-inclination correlation whatsoever, populating two distinct
parts of the Classical Kuiper Belt. We will address this possibility further
ahead. 

We are aware, though, that when considering objects with $i_k<i_k^c$ and
$i_k\geqslant i_k^c$ separately we are also reducing the inclination spans
under analysis, {\it i.e.},  constraining the range of inclination values.  We
saw previously that only about half of the color variability can be
explained by inclination differences. Consequently, the weakening of
correlation values and their significance levels seen when splitting the data
set in two inclination groups, could simply be a consequence of using an
inclination range too narrow to detect any significant trend. To
investigate this possibility we analyze how the mean colors of CKBOs vary with
inclination.


\subsection{Color differences as a function of inclination}
\label{colordiff:sec}

Evidently, if the sample shows a color-inclination trend the mean colors of
CKBOs with $i_k<i_k^c$ must be different from those with  $i_k\geqslant i_k^c$.
That is, they must be statistically incompatible. If the trend was
approximately linear, evidence for color incompatibility would simply vary
smoothly with the number of objects above and below $i_k^c$, as it also depends
on that number. However, if there is a homogenous group of colors below some
$i_k^c$ value then a maximum of color incompatibility between objects above and
below that $i_k^c$ is expected to occur.

Using the Wilcoxon Test, we successively compare the mean colors of 
CKBOs having $i_k<i_k^c$ with those
having $i_k\geqslant i_k^c$, varying $i_{k}^c$ from $3^{\circ}$ to $20^{\circ}$
in increments of $0.5^{\circ}$. Results for each $i_k^c$ are listed in the last
column of Table \ref{rhos_wilcoxon.tab} --- the $W_{SL}$ value is the
significance level in standard deviations of a Gaussian probability
distribution. The mean values are also indicated. 
The significance of these differences peaks at
$i_{k}^c=12.0^{\circ}$, with a value of $6.3\,\sigma$. These results
corroborate the existence of a homogenous set of colors below
$i_{k}\approx12^{\circ}$, as suggested by the analysis in the previous sections.


\subsection{Confidence intervals for critical inclination cutoff}
\label{confintervals:sec}

The finding of the critical inclination cutoff $i_{k}^c=12^{\circ}$ that
separates the red group of CKBOs from the more blue ones, carried out in the
previous sections, assumes that our data set is a representative sample of the
CKBOs. As with the correlation coefficients case, we may use bootstraps to
estimate the confidence interval (error bar) of the best inclination cutoff
$i_k^c$ obtained from the Wilcoxon Tests. We have made 1000 bootstrap
extractions from the data set and for each extraction we have looked for the
$i_k^c$ values that maximized the color differences between objects above and
below it, as done in the previous section. Since in our analysis the $i_k^c$ is
not continuous but discrete (with $0.5^{\circ}$ steps) the bootstrap
distribution is likely to be jagged. To avoid jaggedness a Gaussian noise with
$\sigma=0.25^{\circ}$ was added to the inclinations of each bootstrap
extraction (smooth bootstrap).

\begin{figure}
\epsscale{1.0}
\begin{center}
\plotone{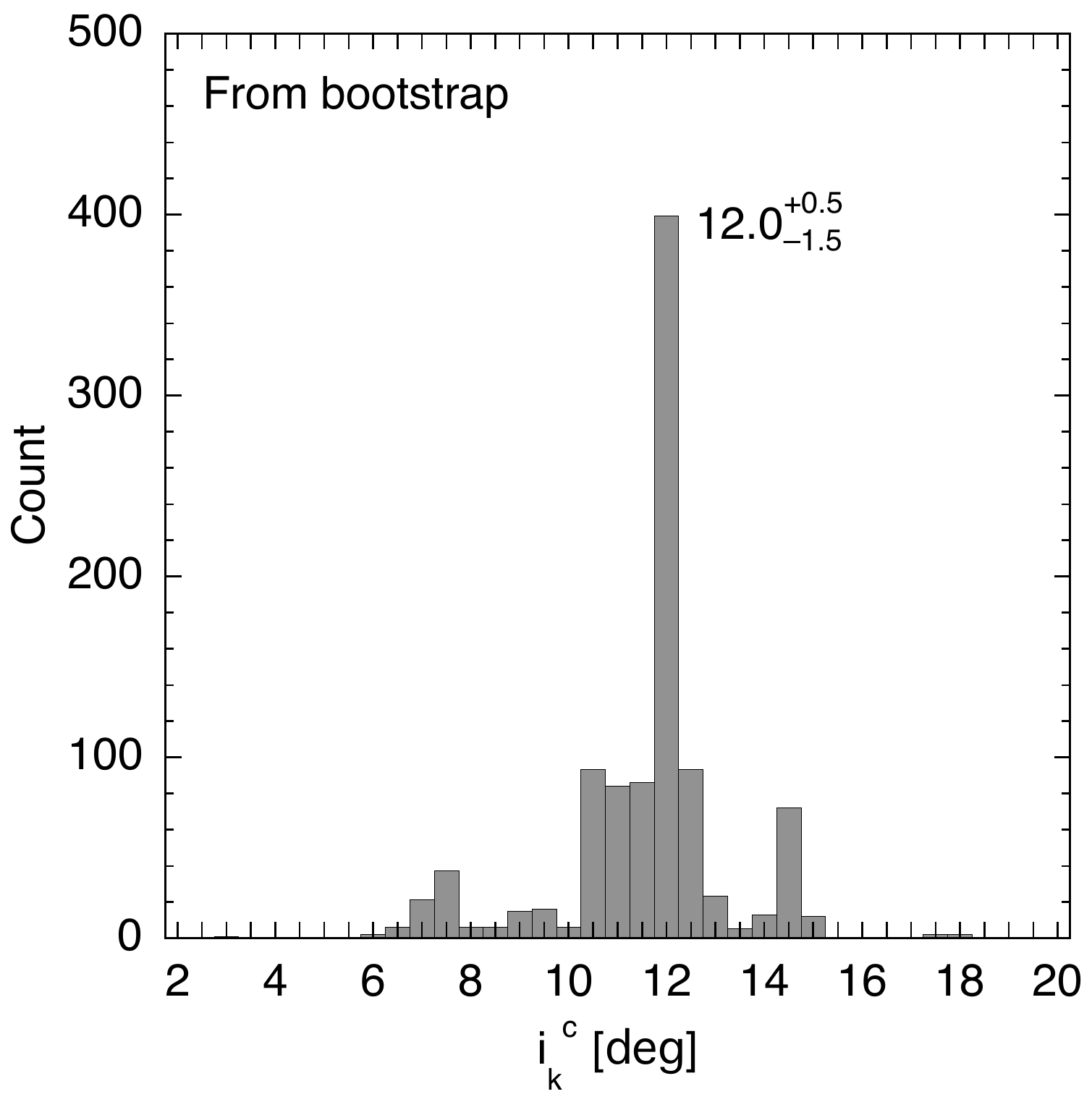}
\caption{Histogram of the probability density distribution of best critical
inclination cutoff $i_k^c$ for maximum color differences (from Wilcoxon Tests),
obtained by bootstrapping our data sample. The maximum color difference is
found at $i_k^c=12.0^{\circ}$$^{+0.5}_{-1.5}$.\label{hist_boots_w_data.fig}
} 
\end{center} 
\end{figure}

The probability density distribution of best critical inclination cutoff
$i_k^c$ for maximum color differences (from Wilcoxon Tests) is shown in
Fig.~\ref{hist_boots_w_data.fig}. The $i_k^c$ that maximizes color differences
is well centered around $12^{\circ}$. Its $1\,\sigma$ confidence interval
($68.3\%$ percentile) is $i_k^c=12.0^{\circ}$$^{+0.5}_{-1.5}$.  The probability
density distribution is not smoothly bell-shaped and two other small solution
spikes are also present: $5.8\%$ probability for
$i_k^c=7.5^{\circ}$$^{+0.0}_{-0.5}$ and $9.7\%$ probability at
$i_k^c=14.5^{\circ}$$^{+0.5}_{-0.5}$.  However, the associated probabilities of
these spikes are low and they do not warrant further attention.

\begin{figure*}
\epsscale{1.1}
\begin{center}
\plotone{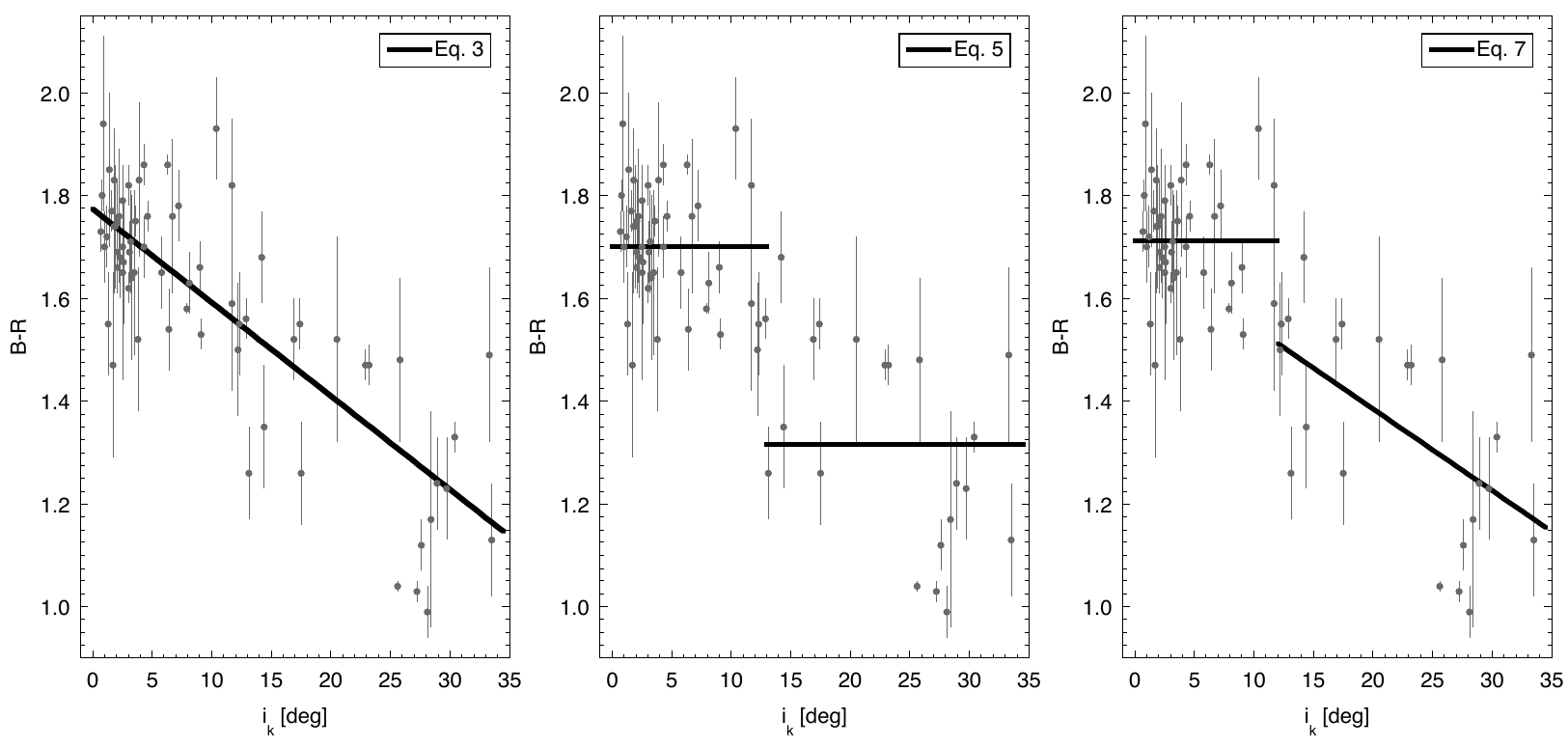}
\caption{Three functional forms studied as possible color-inclination behaviors
of CKBOs. Left: Eq.~\ref{linfit:eq} --- the linear fit; center:
Eq.~\ref{twoconstfit:eq} --- the two-constant stepwise fit with
$i_{k}^c=13^{\circ}$; right: Eq.~\ref{stepfit:eq} --- the
constant-linear stepwise fit with $i_{k}^c=12^{\circ}$. A $\chi^2$ analysis
shows that Eq.~\ref{stepfit:eq} is the best fit. 
\label{ik_br_fits.fig}
} 
\end{center} 
\end{figure*}


\subsection{The color-inclination relation}
\label{bestfit:sec}

Having established that the CKBOs with $i_k\lesssim 12^{\circ}$ constitute a
group of homogeneously red objects, we next examine the variation of $B-R$ at
larger inclinations and the apparent stepwise behavior at the edge of the homogeneously red group.  We consider three different functional forms for $B-R$
color as a function of inclination:
a) linear; b) two-constant stepwise; c) constant-linear stepwise (see Fig.~\ref{ik_br_fits.fig}).

\subsubsection{Linear fit}

Firstly,  we performed a simple linear fit to the data, as:

\begin{equation}
\label{linfitgeneral:eq}
(B-R)=m\,i_k+(B-R)_{o}
\end{equation}

\noindent where $m$ is the linear slope and $(B-R)_{o}$ is the intercept. We
have used a non-weighted Levenberg-Marquardt least-squares fit
\citep{Lev44,Mar63}. 
We chose not to weight the data points using their error bars as those refer to
the precision of each color measurement and not to the expected departure from
the global trend.   
We have obtained the solution:

\begin{equation}
\label{linfit:eq}
(B-R)=-0.0182\,i_k+1.774
\end{equation}

\noindent with a $\chi^2=1.208$ and $df=67$ degrees-of-freedom (see
Fig.~\ref{ik_br_fits.fig}).

\subsubsection{Two-constant stepwise fit}
\label{twoconstfit.sec}

Secondly, since our previous analysis also showed the possibility that CKBOs
consist of two different homogenous groups of objects, neither with any
color-inclination correlation, we fitted the data with a two-constant stepwise
function:

\begin{eqnarray}
\label{twoconstfitgeneral:eq}
\left\{
\begin{array}{ll}
	(B-R)=(B-R)_{o1} & \Leftarrow i_k < i_{k}^c  \\
	(B-R)=(B-R)_{o2} & \Leftarrow i_k \geqslant i_{k}^c
\end{array}
\right.
\end{eqnarray}

\noindent {\it i.e.}, a stepwise function with a constant color value below a
given critical inclination $i_{k}^c$, and with another constant value above
$i_{k}^c$. We have fitted this function to the data iteratively, changing
$i_{k}^c$ from $3^{\circ}$ to $20^{\circ}$ with increments of $0.5^{\circ}$.
For each iteration, $(B-R)_{o1}$ and $(B-R)_{o2}$ are fitted while the critical
$i_{k}^c$ is kept fixed. Table \ref{fits.tab} shows the results obtained for
each $i_{k}^c$ value. The best fit, defined as the one which minimizes
$\chi^2$, is obtained when $i_{k}^c=13.0^{\circ}$:

\begin{eqnarray}
\label{twoconstfit:eq}
\left\{
\begin{array}{ll}
	(B-R)=1.701 & \Leftarrow i_k < 13.0^{\circ}  \\
	(B-R)=1.316 & \Leftarrow i_k \geqslant 13.0^{\circ} 
\end{array}
\right.
\end{eqnarray}

\noindent with $\chi^2=1.349$ and $df=66$ (see
Fig.~\ref{ik_br_fits.fig}).

\subsubsection{Constant-linear stepwise fit}
\label{constlinfit.sec}

Thirdly, we chose to fit a constant-linear stepwise function:

\begin{eqnarray}
\label{stepfitgeneral:eq}
\left\{
\begin{array}{ll}
	(B-R)=(B-R)_{o1} & \Leftarrow i_k < i_{k}^c  \\
	(B-R)=m\,i_k+(B-R)_{o2} & \Leftarrow i_k \geqslant i_{k}^c
\end{array}
\right.
\end{eqnarray}

\noindent where we have a $(B-R)_{o1}$ constant value below some critical
inclination $i_k^c$, and a linear behavior with slope $m$ and intercept
$(B-R)_{o2}$ above $i_k^c$. As for the previous case, we have fitted this
function iteratively changing $i_{k}^c$ from $3^{\circ}$ to $20^{\circ}$ with
increments of $0.5^{\circ}$. For each iteration $(B-R)_{o1}$, $m$, and
$(B-R)_{o2}$ are fitted while $i_{k}^c$ is kept fixed. The fitting results for
each $i_{k}^c$ are shown in Table \ref{fits.tab}.  The best fit, from the
minimum $\chi^2$, is obtained when $i_{k}^c=12.0^{\circ}$:

\begin{eqnarray}
\label{stepfit:eq}
\left\{
\begin{array}{ll}
	(B-R)=1.712 & \Leftarrow i_k < 12.0^{\circ}  \\
	(B-R)=-0.0159\,i_k+1.703 & \Leftarrow i_k \geqslant 12.0^{\circ} 
\end{array}
\right.
\end{eqnarray}

\noindent with $\chi^2=1.100$ and $df=65$ (see
Fig.~\ref{ik_br_fits.fig}). The smallest $\chi^2$ value points to this solution
as the best. 
Note that in our analysis the $\chi^2$ does
not tell us the probability of eventually obtaining a better fit if we had
another sample of CKBOs (with smaller error bars, for example). This last solution is the best relative to the other cases, and validates our findings as discussed in the previous sections.

\subsection{Sharp boundary between groups or overlap?}

We have seen that CKBOs up to $i_k=12\degr$ are homogeneous in terms of
their $B-R$ color (see Fig.~\ref{hist_br_ik_stacked.fig}) and that they are redder than CKBOs with
$i_k>12\degr$.  If this is due to the existence of two independent populations,
then we might expect some mixing of the two at inclinations close to the
boundary. Such mixing might also be expected from dynamical considerations as
cold (low-$i_k$) CKBOs may be pumped to higher inclinations
($10\degr\sim15\degr$) due to interactions with resonances or even with a 
potential ``planetoid'' \citep[{\it e.g.},][]{Kuc+02, LykMuk07, LykMuk08}.

We cannot compute the $\chi^2$ from the superposition of two functions for direct comparison with the 
fits obtained in the previous section. However, we have used the functional core-halo inclination decomposition proposed by
\citet{Ell+05} to consider the possibility that the observed color systematics
(Fig.~\ref{plot_ik_br.fig}) result from the overlap of a red core population (low inclinations)
with a blue halo population (high inclinations). 
Simulations show that this solution fails in the sense that it cannot
reproduce the color jump observed at $12\degr$. The combination of the broad inclination
distribution for the halo objects (see Fig.~20 of \citealt{Ell+05}) and the
large color dispersion we observe for the bluer objects ($1\sigma=0.20$; see
also Fig.~\ref{hist_br_ik_stacked.fig}) results in a very smooth and broad
color distribution at all inclinations, except for $i_k<3\degr$ where red
objects are slightly more abundant. 

Interestingly, the distribution of $B-R$ colors of CKBOs below $i_k=12\degr$ is
remarkably Gaussian, which attests to their color homogeneity. 
The Kolmogorov-Smirnov Test \citep[hereafter KS;][]{Kolmogorov33, Smirnov39} gives a confidence level 
of 99.3\% ($2.7\,\sigma$) that
the colors of $i_k<12\degr$ CKBOs are drawn from a Gaussian distribution with
mean $\mu=1.71$ and standard deviation $\sigma=0.11$ (values calculated from the sample). 
This stands in contrast to the color distribution of objects above 
$i_k=12\degr$ (mean $\mu=1.34$ and
standard deviation $\sigma=0.20$) for which the KS Test  gives a probability of only $22.5\%$ of being derived from a Gaussian. From the KS Test the whole sample of $B-R$ colors of CKBOs has only a $11.4\%$ probability of being Gaussian.

The lack of a break in the color distribution near $5^{\circ}$ stands in sharp contrast to the reportedly bimodal distribution of orbital inclinations \citep{Brown01, Ell+05}. Some models have attempted to relate dynamically cold ($i\lesssim 5^{\circ}$), red KBOs to a primordial trans-Neptunian disk source while dynamically hotter ($i\gtrsim 5^{\circ}$), blue KBOs are supposed to originate by outward scattering  from sources interior to Neptune \citep{Gomes03, Morb+03}. These models make the {\it ad hoc} assumption that hot and cold KBO populations have intrinsically different colors (blue and red, respectively). This assumption is inconsistent with the data. The $B-R$ distribution pays no regard to the reported hot/cold inclination distribution.


\subsection{Double-checks}
\label{checking:sec}

Some of our objects possess large photometric error bars. Under the penalty of too
low sampling we have also looked for the best fitting solution using only the
48 CKBOs with colors having errors $\leqslant 0.10$. The best fitting
solutions are found with equal probability for Eq.~\ref{stepfit:eq} when
$i_{k}^c$ varies from $10.5^{\circ}$ to $12.0^{\circ}$ (with this sample we
have no colors between these two inclination values). When comparing the mean
colors of objects above and below $i_{k}^c$ with this reduced sample we also
find equal incompatibility levels for $i_{k}^c=10.5^{\circ}$ up to
$12.0^{\circ}$, as expected from the previous result. Also, when not discarding the two apparent outliers mentioned in Section \ref{dataset:sec} the correlation values diminish slightly 
but all other results remain identical.

Our sample of CKBOs was selected following a criterion based on the magnitude of the error bars instead of performing an average of all the published measurements for each object as used in the MBOSS sample (see Section \ref{dataset:sec}). 
To check the influence of this criterion we have double-checked our results using the CKBO colors from the MBOSS sample, since some color values were slightly different from those we have used. The outcome is identical to that of Section \ref{bestfit:sec}. 

Lastly, \cite{Glad+08} suggest an orbit classification scheme slightly different from the one used here \citep[by][]{LykMuk07}. The classifications of most KBOs remain unchanged between these two schemes and, not surprisingly, we find that our conclusions are statistically independent of the scheme employed.


\section{Conclusions}

The main goal of this work has been to investigate the
color-inclination trend seen for Classical Kuiper Belt Objects (CKBOs). We have
analyzed a sample of $B-R$ colors of 69 objects, excluding 2
apparent outliers as well as colors with error bars larger than $0.21$.
Objects were classified as CKBOs according to a definition by \cite{LykMuk07}.
Orbital inclinations, denoted $i_k$, were calculated relative to the Kuiper
Belt Plane following \cite{Ell+05}. Our results may be summarized as:

\begin{enumerate}

\item The linear $B-R$ color-inclination correlation of CKBOs measured over the
full range of inclinations from $0\degr$ to $34\degr$ is
$\rho=-0.70^{+0.09}_{-0.07}$, corresponding to a significance level larger than
$8\,\sigma$. This is a strong and highly significant correlation, consistent
with previously published values.

\item In contrast, the $B-R$ colors of CKBOs with inclinations $i_k\leqslant 12.0^{\circ}$$^{+0.5}_{-1.5}$ are
statistically uncorrelated with inclination and are well described by
$B-R=1.71\pm0.11$.  

\item CKBOs with $i_k >12.0^{\circ}$$^{+0.5}_{-1.5}$ show a slight color vs. inclination dependence
following $(B-R)=-0.0159\,i_k+1.703$. The data are also formally
consistent with a constant but bluer color, $B-R = 1.33\pm0.20$, for  $i_k \geqslant 12.5\degr$, and a constant red color $B-R=1.70\pm0.11$ for $i_k
<12.5\degr$. 

\item The data provide no evidence for a break or change in the $B-R$ color
distribution at the boundary between the dynamically
hot and cold populations, purportedly near $i_k \approx 5^{\circ}$.  
In this sense, we find no observational support for the frequently-cited
parity between red CKBOs and the dynamically cold
population. The CKBOs are red up to $i_k \approx 12^{\circ}$ and, therefore 
equally red into the dynamically hot population. 

\end{enumerate}


\acknowledgments

We thank Chadwick Trujillo, Bin Yang, and Rachel Stevenson for comments. 
NP acknowledges funding from the European Social Fund and the Portuguese
Foundation for Science and Technology (FCT, ref.: BPD/ 18729/ 2004). PL and
DJ benefitted from a grant to DJ from the Planetary Astronomy program of the
National Science Foundation. 



\clearpage

\renewcommand{\arraystretch}{0.65}

\begin{deluxetable}{llrrrrl}
\tabletypesize{\scriptsize}
\tablecaption{Data Sample\label{data.tab}}
\tablewidth{0pt}
\tablehead{
\multicolumn{2}{c}{Object} & \colhead{$i_{k}$[$^{\circ}$]\tablenotemark{a}} & \colhead{$i$[$^{\circ}$]\tablenotemark{b}}  & \colhead{$B-R$} & \colhead{$H_R$\tablenotemark{c}} & \colhead{Ref.\tablenotemark{d}}
}
\startdata

                 & 2001QY297 & 0.3 & 1.5 & 1.13 $\pm$ 0.15 & 4.97 $\pm$ 0.23 & Dor+05 \\
                 & 1998WV24 & 0.5 & 1.5 & 1.27 $\pm$ 0.03 & 6.93 $\pm$ 0.01 & TR00 \\
                 & 1998KS65 & 0.7 & 1.2 & 1.73 $\pm$ 0.04 & 6.99 $\pm$ 0.02 & TR03 \\
 (85633)         & 1998KR65 & 0.8 & 1.2 & 1.80 $\pm$ 0.03 & 6.43 $\pm$ 0.03 & TR03 \\
                 & 1999CO153 & 0.9 & 0.8 & 1.94 $\pm$ 0.17 & 6.60 $\pm$ 0.03 & MB(a) \\
 (15760)         & 1992QB1 & 1.0 & 2.2 & 1.70 $\pm$ 0.07 & 6.83 $\pm$ 0.03 & TR00 \\
(134860)         & 2000OJ67 & 1.2 & 1.1 & 1.72 $\pm$ 0.06 & 5.87 $\pm$ 0.07 & Dor+02 \\
 (52747)         & 1998HM151 & 1.3 & 0.5 & 1.55 $\pm$ 0.10 & 7.40 $\pm$ 0.05 & TR03 \\
                 & 2000CL104 & 1.4 & 1.2 & 1.85 $\pm$ 0.15 & 6.87 $\pm$ 0.06 & Boe+02 \\
                 & 2000FS53 & 1.6 & 2.1 & 1.77 $\pm$ 0.04 & 7.17 $\pm$ 0.06 & TR03 \\
 (66652)         & 1999RZ253 & 1.7 & 0.6 & 1.47 $\pm$ 0.18 & 5.42 $\pm$ 0.06 & MB(b) \\
                 & 1994EV3 & 1.8 & 1.7 & 1.74 $\pm$ 0.13 & 7.53 $\pm$ 0.09 & Boe+02 \\
 (66452)         & 1999OF4 & 1.8 & 2.7 & 1.83 $\pm$ 0.10 & 6.10 $\pm$ 0.09 & Pei+04 \\
                 & 1999OM4 & 1.9 & 2.1 & 1.74 $\pm$ 0.12 & 7.43 $\pm$ 0.06 & Boe+02 \\
 (79360)         & 1997CS29 & 2.1 & 2.2 & 1.69 $\pm$ 0.08 & 4.91 $\pm$ 0.11 & TR98 \\
(119951)         & 2002KX14 & 2.1 & 0.4 & 1.66 $\pm$ 0.04 & 4.25 $\pm$ 0.03 & TRC07 \\
                 & 2003GH55 & 2.1 & 1.1 & 1.75 $\pm$ 0.08 & 5.90 $\pm$ 0.05 & JPH07 \\
                 & 1998KG62 & 2.2 & 0.8 & 1.76 $\pm$ 0.13 & 6.92 $\pm$ 0.08 & Boe+02 \\
 (19255)         & 1994VK8 & 2.3 & 1.5 & 1.68 $\pm$ 0.07 & 6.86 $\pm$ 0.42 & TR00 \\
                 & 1999OJ4 & 2.3 & 4.0 & 1.68 $\pm$ 0.08 & 6.71 $\pm$ 0.06 & Pei+04 \\
                 & 1994ES2 & 2.5 & 1.1 & 1.65 $\pm$ 0.21 & 7.52 $\pm$ 0.12 & MB(c) \\
                 & 1998WX24 & 2.5 & 0.9 & 1.79 $\pm$ 0.07 & 6.09 $\pm$ 0.04 & TR00 \\
 (60454)         & 2000CH105 & 2.5 & 1.2 & 1.70 $\pm$ 0.08 & 6.20 $\pm$ 0.05 & Pei+04 \\
 (58534) Logos   & 1997CQ29 & 2.6 & 2.9 & 1.67 $\pm$ 0.12 & 6.70 $\pm$ 0.02 & Bar+01 \\
                 & 1996TK66 & 3.0 & 3.3 & 1.62 $\pm$ 0.03 & 6.12 $\pm$ 0.03 & TR00 \\
 (24978)         & 1998HJ151 & 3.0 & 2.4 & 1.82 $\pm$ 0.04 & 6.96 $\pm$ 0.02 & TR03 \\
(137294)         & 1999RE215 & 3.1 & 1.4 & 1.69 $\pm$ 0.06 & 6.45 $\pm$ 0.17 & Boe+02 \\
 (33001)         & 1997CU29 & 3.2 & 1.5 & 1.71 $\pm$ 0.10 & 6.12 $\pm$ 0.06 & Dor+01 \\
                 & 2001QD298 & 3.3 & 5.0 & 1.64 $\pm$ 0.16 & 4.48 $\pm$ 0.08 & Dor+05 \\
(148780)         & 2001UQ18 & 3.5 & 5.2 & 1.65 $\pm$ 0.16 & 5.82 $\pm$ 0.21 & Dor+05 \\
 (16684)         & 1994JQ1 & 3.6 & 3.7 & 1.75 $\pm$ 0.03 & 6.51 $\pm$ 0.03 & TR03 \\
                 & 2000CL105 & 3.8 & 4.2 & 1.52 $\pm$ 0.14 & 6.76 $\pm$ 0.06 & MB(a) \\
                 & 1999OE4 & 3.9 & 2.2 & 1.83 $\pm$ 0.15 & 6.76 $\pm$ 0.17 & Pei+04 \\
                 & 1999HS11 & 4.3 & 2.6 & 1.86 $\pm$ 0.04 & 6.16 $\pm$ 0.03 & TR03 \\
                 & 1999HV11 & 4.3 & 3.2 & 1.70 $\pm$ 0.06 & 6.88 $\pm$ 0.03 & TR03 \\
                 & 2000CN105 & 4.6 & 3.4 & 1.76 $\pm$ 0.03 & 5.21 $\pm$ 0.05 & JPH07 \\
                 & 1999RX214 & 5.8 & 4.8 & 1.65 $\pm$ 0.07 & 6.32 $\pm$ 0.05 & Pei+04 \\
                 & 1997CV29 & 6.3 & 8.0 & 1.86 $\pm$ 0.02 & 7.06 $\pm$ 0.01 & TR03 \\
(138537)         & 2000OK67 & 6.4 & 4.9 & 1.54 $\pm$ 0.08 & 5.92 $\pm$ 0.07 & Dor+02 \\
                 & 1999GS46 & 6.7 & 5.2 & 1.76 $\pm$ 0.15 & 6.23 $\pm$ 0.02 & MB(a) \\
                 & 1996TS66 & 7.2 & 7.4 & 1.78 $\pm$ 0.07 & 5.74 $\pm$ 0.08 & TR98 \\
 (50000) Quaoar  & 2002LM60 & 7.9 & 8.0 & 1.58 $\pm$ 0.01 & 2.10 $\pm$ 0.01 & TRC03 \\
 (79983)         & 1999DF9 & 8.1 & 9.8 & 1.63 $\pm$ 0.06 & 5.62 $\pm$ 0.07 & Dor+02 \\
                 & 1993FW & 9.0 & 7.7 & 1.66 $\pm$ 0.05 & 6.46 $\pm$ 0.01 & TR03 \\
                 & 1998FS144 & 9.1 & 9.8 & 1.53 $\pm$ 0.03 & 6.60 $\pm$ 0.02 & TR03 \\
                 & 1999CB119 & 10.4 & 8.7 & 1.93 $\pm$ 0.10 & 6.57 $\pm$ 0.05 & Pei+04 \\
                 & 1999JD132 & 11.7 & 10.5 & 1.59 $\pm$ 0.17 & 6.00 $\pm$ 0.02 & MB(a) \\
                 & 2001KA77 & 11.7 & 11.9 & 1.82 $\pm$ 0.13 & 4.95 $\pm$ 0.04 & Pei+04 \\
                 & 2002GJ32 & 12.2 & 11.6 & 1.50 $\pm$ 0.13 & 5.48 $\pm$ 0.15 & Dor+05 \\
                 & 1997RT5 & 12.3 & 12.7 & 1.55 $\pm$ 0.10 & 7.46 $\pm$ 0.07 & Boe+02 \\
 (19521) Chaos   & 1998WH24 & 12.9 & 12.1 & 1.56 $\pm$ 0.04 & 4.32 $\pm$ 0.01 & TR00 \\
                 & 1999RY214 & 13.1 & 13.7 & 1.26 $\pm$ 0.09 & 6.96 $\pm$ 0.04 & Pei+04 \\
                 & 1997QH4 & 14.2 & 13.2 & 1.68 $\pm$ 0.09 & 6.77 $\pm$ 0.04 & TR00 \\
                 & 1999CQ133 & 14.4 & 13.3 & 1.35 $\pm$ 0.12 & 6.68 $\pm$ 0.05 & MB(a) \\
 (20000) Varuna  & 2000WR106 & 16.9 & 17.2 & 1.52 $\pm$ 0.08 & 3.36 $\pm$ 0.05 & TR03 \\
                 & 2000KK4 & 17.4 & 19.1 & 1.55 $\pm$ 0.05 & 5.82 $\pm$ 0.02 & TR03 \\
 (15883)         & 1997CR29 & 17.5 & 19.2 & 1.26 $\pm$ 0.10 & 6.95 $\pm$ 0.08 & Dor+01 \\
                 & 2000CO105 & 20.5 & 19.3 & 1.52 $\pm$ 0.20 & 5.67 $\pm$ 0.18 & MB(a) \\
 (55565)         & 2002AW197 & 22.9 & 24.4 & 1.47 $\pm$ 0.03 & 3.07 $\pm$ 0.02 & TRC07 \\
 (90568)         & 2004GV9 & 23.2 & 21.9 & 1.47 $\pm$ 0.04 & 3.62 $\pm$ 0.03 & TRC07 \\
 (24835)         & 1995SM55 & 25.6 & 27.1 & 1.04 $\pm$ 0.01 & 4.15 $\pm$ 0.01 & TR03 \\
                 & 2002GH32 & 25.8 & 26.6 & 1.48 $\pm$ 0.16 & 6.05 $\pm$ 0.28 & Dor+05 \\
 (55636)         & 2002TX300 & 27.2 & 25.9 & 1.03 $\pm$ 0.02 & 3.11 $\pm$ 0.01 & TRC03 \\
 (19308)         & 1996TO66 & 27.6 & 27.5 & 1.12 $\pm$ 0.05 & 4.38 $\pm$ 0.05 & TR98 \\
                 & 2003UZ117 & 28.1 & 27.5 & 0.99 $\pm$ 0.05 & 4.85 $\pm$ 0.05 & TRC07 \\
                 & 2000CG105 & 28.4 & 28.0 & 1.17 $\pm$ 0.21 & 6.77 $\pm$ 0.16 & MB(a) \\
                 & 2001QC298 & 28.9 & 30.6 & 1.24 $\pm$ 0.09 & 6.39 $\pm$ 0.05 & JPH07 \\
                 & 1998WT31 & 29.7 & 28.7 & 1.23 $\pm$ 0.10 & 7.40 $\pm$ 0.04 & Pei+04 \\
(136472)         & 2005FY9 & 30.4 & 29.0 & 1.33 $\pm$ 0.03 & -0.38 $\pm$ 0.05 & JPH07 \\
                 & 1996RQ20 & 33.3 & 31.7 & 1.49 $\pm$ 0.17 & 6.89 $\pm$ 0.10 & MB(d) \\
                 & 2002PP149 & 33.5 & 34.7 & 1.13 $\pm$ 0.11 & 7.24 $\pm$ 0.05 & JPH07 \\

\enddata

\tablenotetext{a}{Orbital inclination relative to the Kuiper Belt Plane}
\tablenotetext{b}{Orbital inclination relative to the Ecliptic}
\tablenotetext{c}{Absolute $R$-magnitude}
\tablenotetext{d}{References: TRC07, http://www.physics.nau.edu/$\sim$tegler/research/survey.htm; TRC03, \cite{TRC03}; TR00, \cite{TegRom00}; TR98, \cite{TegRom98};
Boe+02, \cite{Boe+02}; Pei+04, \cite{Peix+04}; JPH07, \cite{JewPeiHsi07}; Dor+05, \cite{Dor+05}; Dor+02, \cite{Dor+02}; Dor+01, \cite{Dor+01}; Bar+01, \cite{Bar+01};
MB, MBOSS compilation \citep{HaiDel02} -- (a) \cite{TruBro02} -- (b) \cite{Del+01, Dor+01, McB+03} -- (c) \cite{Green+97, LuuJew96} -- (d) \cite{TegRom98, RomTeg99, Boe+01, Del+01, JewLuu01}}

\end{deluxetable}

\renewcommand{\arraystretch}{1}


\clearpage

\begin{deluxetable}{rrrrrrrrrrrrr}
\setlength{\tabcolsep}{0.03in}
\tabletypesize{\scriptsize}
\tablecaption{Correlations and Wilcoxon Tests for Consecutive Inclination Cutoffs \label{rhos_wilcoxon.tab}}
\tablewidth{0pt}
\tablehead{
\colhead{} & \colhead{} & \multicolumn{4}{c}{Objects w/ $i_k<i_{k}^c$} & \colhead{} & \multicolumn{4}{c}{Objects w/ $i_k\geqslant i_{k}^c$} & & \\
\cline{3-6} \cline{8-11} \\ 
\colhead{$i_{k}^c$[$^{\circ}$]\tablenotemark{a}} & &
\colhead{$N_l$\tablenotemark{b}} & \colhead{$\mu_l$\tablenotemark{c}} &
\colhead{$\rho_l$$^{+\sigma}_{-\sigma}$\tablenotemark{d}} &
\colhead{$SL_l$\tablenotemark{e}} & \colhead{} &
\colhead{$N_h$\tablenotemark{b}} & \colhead{$\mu_h$\tablenotemark{c}} &
\colhead{$\rho_h$$^{+\sigma}_{-\sigma}$\tablenotemark{d}} &
\colhead{$SL_h$\tablenotemark{e}} & \colhead{} & $W_{SL}$\tablenotemark{f} 
}
\startdata

 3.0 & &  22 & 1.721 & $ -0.34 ^{+ 0.21 }_{ -0.18 } $ & $1.6$  & &  47 & 1.528 & $ -0.79 ^{+ 0.06 }_{ -0.04 } $ & $>8$ & & $3.5$     \\
 3.5 & &  27 & 1.717 & $ -0.35 ^{+ 0.21 }_{ -0.18 } $ & $1.8$  & &  42 & 1.508 & $ -0.81 ^{+ 0.05 }_{ -0.04 } $ & $>8$ & & $3.8$     \\
 4.0 & &  31 & 1.713 & $ -0.28 ^{+ 0.21 }_{ -0.19 } $ & $1.5$  & &  38 & 1.489 & $ -0.84 ^{+ 0.04 }_{ -0.04 } $ & $>8$ & & $4.1$     \\
 4.5 & &  33 & 1.717 & $ -0.17 ^{+ 0.20 }_{ -0.19 } $ & $1.0$  & &  36 & 1.473 & $ -0.82 ^{+ 0.05 }_{ -0.04 } $ & $>8$ & & $4.6$     \\
 5.0 & &  34 & 1.718 & $ -0.13 ^{+ 0.21 }_{ -0.20 } $ & $0.7$  & &  35 & 1.465 & $ -0.81 ^{+ 0.05 }_{ -0.04 } $ & $>8$ & & $4.8$     \\
 5.5 & &  34 & 1.718 & $ -0.13 ^{+ 0.19 }_{ -0.18 } $ & $0.7$  & &  35 & 1.465 & $ -0.81 ^{+ 0.05 }_{ -0.04 } $ & $>8$ & & $4.8$     \\
 6.0 & &  35 & 1.716 & $ -0.17 ^{+ 0.20 }_{ -0.18 } $ & $1.0$  & &  34 & 1.459 & $ -0.81 ^{+ 0.05 }_{ -0.04 } $ & $>8$ & & $4.8$     \\
 6.5 & &  37 & 1.715 & $ -0.15 ^{+ 0.21 }_{ -0.20 } $ & $0.9$  & &  32 & 1.444 & $ -0.81 ^{+ 0.06 }_{ -0.05 } $ & $>8$ & & $5.0$     \\
 7.0 & &  38 & 1.717 & $ -0.11 ^{+ 0.19 }_{ -0.18 } $ & $0.7$  & &  31 & 1.434 & $ -0.80 ^{+ 0.06 }_{ -0.05 } $ &  5.4 & & $5.3$     \\
 7.5 & &  39 & 1.718 & $ -0.07 ^{+ 0.19 }_{ -0.18 } $ & $0.4$  & &  30 & 1.423 & $ -0.78 ^{+ 0.07 }_{ -0.05 } $ &  5.2 & & $5.6$     \\
 8.0 & &  40 & 1.715 & $ -0.12 ^{+ 0.19 }_{ -0.18 } $ & $0.7$  & &  29 & 1.417 & $ -0.78 ^{+ 0.07 }_{ -0.06 } $ &  5.0 & & $5.5$     \\
 8.5 & &  41 & 1.713 & $ -0.17 ^{+ 0.18 }_{ -0.17 } $ & $1.0$  & &  28 & 1.410 & $ -0.76 ^{+ 0.09 }_{ -0.07 } $ &  4.7 & & $5.5$     \\
 9.0 & &  41 & 1.713 & $ -0.17 ^{+ 0.17 }_{ -0.16 } $ & $1.0$  & &  28 & 1.410 & $ -0.76 ^{+ 0.08 }_{ -0.06 } $ &  4.7 & & $5.5$     \\
 9.5 & &  43 & 1.707 & $ -0.24 ^{+ 0.16 }_{ -0.15 } $ & $1.6$  & &  26 & 1.395 & $ -0.74 ^{+ 0.12 }_{ -0.09 } $ &  4.3 & & $5.4$     \\
10.0 & &  43 &1.707 & $ -0.24 ^{+ 0.18 }_{ -0.17 } $ & $1.6$  & &  26 & 1.395 & $ -0.74 ^{+ 0.12 }_{ -0.09 } $ &  4.3 & & $5.4$     \\
10.5 & &  44 &1.712 & $ -0.17 ^{+ 0.18 }_{ -0.17 } $ & $1.1$  & &  25 & 1.374 & $ -0.70 ^{+ 0.12 }_{ -0.09 } $ &  3.9 & & $5.9$     \\
11.0 & &  44 &1.712 & $ -0.17 ^{+ 0.18 }_{ -0.17 } $ & $1.1$  & &  25 & 1.374 & $ -0.70 ^{+ 0.12 }_{ -0.09 } $ &  3.9 & & $5.9$     \\
11.5 & &  44 &1.712 & $ -0.17 ^{+ 0.18 }_{ -0.17 } $ & $1.1$  & &  25 & 1.374 & $ -0.70 ^{+ 0.12 }_{ -0.09 } $ &  3.9 & & $5.9$     \\
\hline
12.0 & &  46 &1.712 & $ -0.15 ^{+ 0.18 }_{ -0.17 } $ & $1.0$  & &  23 & 1.345 & $ -0.62 ^{+ 0.14 }_{ -0.11 } $ &  3.2 & & $6.3$     \\
\hline
12.5 & &  48 &1.704 & $ -0.24 ^{+ 0.17 }_{ -0.16 } $ & $1.6$  & &  21 & 1.328 & $ -0.55 ^{+ 0.18 }_{ -0.14 } $ &  2.6 & & $6.1$     \\
13.0 & &  49 &1.701 & $ -0.27 ^{+ 0.17 }_{ -0.15 } $ & $1.9$  & &  20 & 1.317 & $ -0.48 ^{+ 0.18 }_{ -0.15 } $ &  2.2 & & $6.0$     \\
13.5 & &  50 &1.692 & $ -0.32 ^{+ 0.17 }_{ -0.15 } $ & $2.2$  & &  19 & 1.319 & $ -0.54 ^{+ 0.23 }_{ -0.17 } $ &  2.4 & & $5.8$     \\
14.0 & &  50 &1.692 & $ -0.32 ^{+ 0.17 }_{ -0.15 } $ & $2.2$  & &  19 & 1.319 & $ -0.54 ^{+ 0.22 }_{ -0.16 } $ &  2.4 & & $5.8$     \\
14.5 & &  52 &1.686 & $ -0.36 ^{+ 0.15 }_{ -0.14 } $ & $2.6$  & &  17 & 1.296 & $ -0.45 ^{+ 0.28 }_{ -0.21 } $ &  1.8 & & $5.7$     \\
15.0 & &  52 &1.686 & $ -0.36 ^{+ 0.15 }_{ -0.14 } $ & $2.6$  & &  17 & 1.296 & $ -0.45 ^{+ 0.27 }_{ -0.21 } $ &  1.8 & & $5.7$     \\
15.5 & &  52 &1.686 & $ -0.36 ^{+ 0.15 }_{ -0.14 } $ & $2.6$  & &  17 & 1.296 & $ -0.45 ^{+ 0.26 }_{ -0.20 } $ &  1.8 & & $5.7$     \\
16.0 & &  52 &1.686 & $ -0.36 ^{+ 0.15 }_{ -0.14 } $ & $2.6$  & &  17 & 1.296 & $ -0.45 ^{+ 0.27 }_{ -0.21 } $ &  1.8 & & $5.7$     \\
16.5 & &  52 &1.686 & $ -0.36 ^{+ 0.15 }_{ -0.14 } $ & $2.6$  & &  17 & 1.296 & $ -0.45 ^{+ 0.26 }_{ -0.20 } $ &  1.8 & & $5.7$     \\
17.0 & &  53 &1.682 & $ -0.39 ^{+ 0.15 }_{ -0.13 } $ & $2.9$  & &  16 & 1.282 & $ -0.35 ^{+ 0.29 }_{ -0.24 } $ &  1.3 & & $5.7$     \\
17.5 & &  54 &1.680 & $ -0.41 ^{+ 0.14 }_{ -0.12 } $ & $3.0$  & &  15 & 1.265 & $ -0.21 ^{+ 0.32 }_{ -0.28 } $ &  0.8 & & $5.6$     \\
18.0 & &  55 &1.672 & $ -0.44 ^{+ 0.14 }_{ -0.12 } $ & $3.4$  & &  14 & 1.265 & $ -0.19 ^{+ 0.37 }_{ -0.32 } $ &  0.7 & & $5.4$     \\
18.5 & &  55 &1.672 & $ -0.44 ^{+ 0.15 }_{ -0.12 } $ & $3.4$  & &  14 & 1.265 & $ -0.19 ^{+ 0.35 }_{ -0.31 } $ &  0.7 & & $5.4$     \\
19.0 & &  55 &1.672 & $ -0.44 ^{+ 0.15 }_{ -0.13 } $ & $3.4$  & &  14 & 1.265 & $ -0.19 ^{+ 0.36 }_{ -0.32 } $ &  0.7 & & $5.4$     \\
19.5 & &  55 &1.672 & $ -0.44 ^{+ 0.14 }_{ -0.12 } $ & $3.4$  & &  14 & 1.265 & $ -0.19 ^{+ 0.35 }_{ -0.31 } $ &  0.7 & & $5.4$     \\
20.0 & &  55 &1.672 & $ -0.44 ^{+ 0.14 }_{ -0.12 } $ & $3.4$  & &  14 & 1.265 & $ -0.19 ^{+ 0.37 }_{ -0.32 } $ &  0.7 & & $5.4$     \\
\enddata
\tablenotetext{a}{Orbital inclination cutoff}
\tablenotetext{b}{Number of objects below ($_l$) and above ($_h$) the cutoff $i_{k}^c$}
\tablenotetext{c}{Mean $B-R$ of objects below ($_l$) and above ($_h$) $i_{k}^c$}
\tablenotetext{d}{$B-R$ vs. $i_k$ correlation for objects below ($_l$) and above ($_h$) $i_{k}^c$}
\tablenotetext{e}{Significance level of the correlation}
\tablenotetext{f}{Significance level of the Wilcoxon Test for color difference between objects below and above $i_{k}^c$}

\end{deluxetable}


\clearpage

\begin{deluxetable}{rrcccrcccc}
\setlength{\tabcolsep}{0.03in}
\tabletypesize{\scriptsize}
\tablecaption{Fit Results for Consecutive Inclination Cutoffs \label{fits.tab}}
\tablewidth{0pt}
\tablehead{
\colhead{} & \colhead{} & \multicolumn{3}{c}{Two-constant stepwise fit\tablenotemark{b}} & \colhead{} & \multicolumn{4}{c}{Constant-linear stepwise fit\tablenotemark{c}} \\
\cline{3-5} \cline{7-10} \\ 
\colhead{$i_{k}^c$[$^{\circ}$]\tablenotemark{a}} & & \colhead{$(B-R)_{o1}$} & \colhead{$(B-R)_{o2}$} & \colhead{$\chi^2$} & \colhead{} & \colhead{$(B-R)_{o1}$} & \colhead{$m$} & \colhead{$(B-R)_{o2}$} & \colhead{$\chi^2$}
}

\startdata
 3.0 & & 1.721 & 1.528 & 2.892 & & 1.721 & -0.0194 & 1.801 & 1.197 \\ 
 3.5 & & 1.717 & 1.508 & 2.737 & & 1.717 & -0.0201 & 1.817 & 1.188 \\ 
 4.0 & & 1.713 & 1.489 & 2.597 & & 1.713 & -0.0210 & 1.837 & 1.174 \\ 
 4.5 & & 1.717 & 1.473 & 2.427 & & 1.717 & -0.0207 & 1.830 & 1.180 \\ 
 5.0 & & 1.718 & 1.465 & 2.344 & & 1.718 & -0.0206 & 1.827 & 1.181 \\ 
 5.5 & & 1.718 & 1.465 & 2.344 & & 1.718 & -0.0206 & 1.827 & 1.181 \\ 
 6.0 & & 1.716 & 1.459 & 2.314 & & 1.716 & -0.0208 & 1.834 & 1.182 \\ 
 6.5 & & 1.715 & 1.444 & 2.191 & & 1.715 & -0.0208 & 1.834 & 1.183 \\ 
 7.0 & & 1.717 & 1.434 & 2.090 & & 1.717 & -0.0205 & 1.825 & 1.180 \\ 
 7.5 & & 1.718 & 1.423 & 1.971 & & 1.718 & -0.0198 & 1.809 & 1.173 \\ 
 8.0 & & 1.715 & 1.417 & 1.964 & & 1.715 & -0.0203 & 1.821 & 1.185 \\ 
 8.5 & & 1.713 & 1.410 & 1.924 & & 1.713 & -0.0205 & 1.825 & 1.192 \\ 
 9.0 & & 1.713 & 1.410 & 1.924 & & 1.713 & -0.0205 & 1.825 & 1.192 \\ 
 9.5 & & 1.707 & 1.395 & 1.876 & & 1.707 & -0.0212 & 1.844 & 1.213 \\ 
10.0 & & 1.707 & 1.395 & 1.876 & & 1.707 & -0.0212 & 1.844 & 1.213 \\ 
10.5 & & 1.712 & 1.374 & 1.627 & & 1.712 & -0.0187 & 1.777 & 1.155 \\ 
11.0 & & 1.712 & 1.374 & 1.627 & & 1.712 & -0.0187 & 1.777 & 1.155 \\ 
11.5 & & 1.712 & 1.374 & 1.627 & & 1.712 & -0.0187 & 1.777 & 1.155 \\ 
\cline{1-2} \cline{7-10}
12.0 & & 1.712 & 1.345 & 1.389 & & 1.712 & -0.0159 & 1.703 & 1.100 \\ 
\cline{1-2} \cline{7-10}
12.5 & & 1.704 & 1.328 & 1.385 & & 1.704 & -0.0155 & 1.691 & 1.166 \\ 

\cline{1-5}

13.0 & & 1.701 & 1.316 & 1.349 & & 1.701 & -0.0145 & 1.665 & 1.181 \\ 
\cline{1-5}

13.5 & & 1.692 & 1.319 & 1.537 & & 1.692 & -0.0182 & 1.765 & 1.315 \\ 
14.0 & & 1.692 & 1.319 & 1.537 & & 1.692 & -0.0182 & 1.765 & 1.315 \\ 
14.5 & & 1.686 & 1.296 & 1.512 & & 1.686 & -0.0176 & 1.748 & 1.376 \\ 
15.0 & & 1.686 & 1.296 & 1.512 & & 1.686 & -0.0176 & 1.748 & 1.376 \\ 
15.5 & & 1.686 & 1.296 & 1.512 & & 1.686 & -0.0176 & 1.748 & 1.376 \\ 
16.0 & & 1.686 & 1.296 & 1.512 & & 1.686 & -0.0176 & 1.748 & 1.376 \\ 
16.5 & & 1.686 & 1.296 & 1.512 & & 1.686 & -0.0176 & 1.748 & 1.376 \\ 
17.0 & & 1.682 & 1.282 & 1.486 & & 1.682 & -0.0158 & 1.696 & 1.397 \\ 
17.5 & & 1.680 & 1.265 & 1.427 & & 1.680 & -0.0114 & 1.570 & 1.391 \\ 
18.0 & & 1.672 & 1.265 & 1.600 & & 1.672 & -0.0175 & 1.746 & 1.544 \\ 
18.5 & & 1.672 & 1.265 & 1.600 & & 1.672 & -0.0175 & 1.746 & 1.544 \\ 
19.0 & & 1.672 & 1.265 & 1.600 & & 1.672 & -0.0175 & 1.746 & 1.544 \\ 
19.5 & & 1.672 & 1.265 & 1.600 & & 1.672 & -0.0175 & 1.746 & 1.544 \\ 
20.0 & & 1.672 & 1.265 & 1.600 & & 1.672 & -0.0175 & 1.746 & 1.544 \\
\enddata
\tablenotetext{a}{Critical orbital inclination value, {\it i.e.} location of stepwise behavior.}
\tablenotetext{b}{See \S\ref{twoconstfit.sec} and Eq.~\ref{twoconstfitgeneral:eq}.}
\tablenotetext{c}{See \S\ref{constlinfit.sec} and Eq.~\ref{stepfitgeneral:eq}.}

\end{deluxetable}



\begin{thebibliography}{43}
\expandafter\ifx\csname natexlab\endcsname\relax\def\natexlab#1{#1}\fi

\bibitem[{{Barucci} {et~al.}(2001){Barucci}, {Fulchignoni}, {Birlan},
  {Doressoundiram}, {Romon}, \& {Boehnhardt}}]{Bar+01}
{Barucci}, M.~A., {Fulchignoni}, M., {Birlan}, M., {Doressoundiram}, A.,
  {Romon}, J., \& {Boehnhardt}, H. 2001, Astron. Astrophys., 371, 1150

\bibitem[{{Boehnhardt} {et~al.}(2002){Boehnhardt}, {Delsanti}, {Barucci},
  {Hainaut}, {Doressoundiram}, {Lazzarin}, {Barrera}, {de Bergh}, {Birkle},
  {Dotto}, {Meech}, {Ortiz}, {Romon}, {Sekiguchi}, {Thomas}, {Tozzi},
  {Watanabe}, \& {West}}]{Boe+02}
{Boehnhardt}, H., {et~al.} 2002, Astron. Astrophys., 395, 297

\bibitem[{{Boehnhardt} {et~al.}(2001){Boehnhardt}, {Tozzi}, {Birkle},
  {Hainaut}, {Sekiguchi}, {Vair}, {Watanabe}, {Rupprecht}, \& {The FORS
  Instrument Team}}]{Boe+01}
---. 2001, Astron. Astrophys., 378, 653

\bibitem[{{Brown}(2001)}]{Brown01}
{Brown}, M.~E. 2001, Astrophys. J., 121, 2804

\bibitem[{{Delsanti} {et~al.}(2004){Delsanti}, {Hainaut}, {Jourdeuil}, {Meech},
  {Boehnhardt}, \& {Barrera}}]{Del+04}
{Delsanti}, A., {Hainaut}, O., {Jourdeuil}, E., {Meech}, K., {Boehnhardt}, H.,
  \& {Barrera}, L. 2004, Astron. Astrophys., 417, 1145

\bibitem[{{Delsanti} {et~al.}(2001){Delsanti}, {Boehnhardt}, {Barrera},
  {Meech}, {Sekiguchi}, \& {Hainaut}}]{Del+01}
{Delsanti}, A.~C., {Boehnhardt}, H., {Barrera}, L., {Meech}, K.~J.,
  {Sekiguchi}, T., \& {Hainaut}, O.~R. 2001, Astron. Astrophys., 380, 347

\bibitem[{{Doressoundiram} {et~al.}(2001){Doressoundiram}, {Barucci}, {Romon},
  \& {Veillet}}]{Dor+01}
{Doressoundiram}, A., {Barucci}, M., {Romon}, J., \& {Veillet}, C. 2001,
  Icarus, 154, 277

\bibitem[{{Doressoundiram} {et~al.}(2002){Doressoundiram}, {Peixinho}, {de
  Bergh}, {Fornasier}, {Th{\' e}bault}, {Barucci}, \& {Veillet}}]{Dor+02}
{Doressoundiram}, A., {Peixinho}, N., {de Bergh}, C., {Fornasier}, S., {Th{\'
  e}bault}, P., {Barucci}, M.~A., \& {Veillet}, C. 2002, Astron. J., 124, 2279

\bibitem[{{Doressoundiram} {et~al.}(2005){Doressoundiram}, {Peixinho},
  {Doucet}, {Mousis}, {Barucci}, {Petit}, \& {Veillet}}]{Dor+05}
{Doressoundiram}, A., {Peixinho}, N., {Doucet}, C., {Mousis}, O., {Barucci},
  M.~A., {Petit}, J.~M., \& {Veillet}, C. 2005, Icarus, 174, 90

\bibitem[{Doressoundiram} {et~al.}(2008)]{Dor+08} Doressoundiram, 
A., Boehnhardt, H., Tegler, S.~C., 
\& Trujillo, C.\ 2008, The Solar System Beyond Neptune, 91 

\bibitem[{{Efron} \& {Tibshirani}(1993)}]{EfrTib93}
{Efron}, B., \& {Tibshirani}, R.~J. 1993, {An Introduction to the Bootstrap}
  (Chapman \& Hall/CRC)

\bibitem[{{Elliot} {et~al.}(2005){Elliot}, {Kern}, {Clancy}, {Gulbis},
  {Millis}, {Buie}, {Wasserman}, {Chiang}, {Jordan}, {Trilling}, \&
  {Meech}}]{Ell+05}
{Elliot}, J.~L., {et~al.} 2005, Astron. J., 129, 1117

\bibitem[{{Gil-Hutton}(2002)}]{GilH02}
{Gil-Hutton}, R. 2002, Planet. Space Sci., 50, 57

\bibitem[Gladman et al.(2008)]{Glad+08} Gladman, B., Marsden, 
B.~G., \& Vanlaerhoven, C.\ 2008, The Solar System Beyond Neptune, 43 

\bibitem[{{Gomes}(2003)}]{Gomes03}
{Gomes}, R.~S. 2003, Icarus, 161, 404

\bibitem[{{Green} {et~al.}(1997){Green}, {McBride}, {O Ceallaigh},
  {Fitzsimmons}, {Williams}, \& {Irwin}}]{Green+97}
{Green}, S.~F., {McBride}, N., {O Ceallaigh}, D.~P., {Fitzsimmons}, A.,
  {Williams}, I.~P., \& {Irwin}, M.~J. 1997, Mon. Not. R. Astron. Soc., 290,
  186

\bibitem[{{Gulbis} {et~al.}(2006){Gulbis}, {Elliot}, \& {Kane}}]{Gul+06}
{Gulbis}, A.~A.~S., {Elliot}, J.~L., \& {Kane}, J.~F. 2006, Icarus, 183, 168

\bibitem[{{Hainaut} \& {Delsanti}(2002)}]{HaiDel02}
{Hainaut}, O.~R., \& {Delsanti}, A.~C. 2002, Astron. Astrophys., 389, 641

\bibitem[{{Hartmann} {et~al.}(1990){Hartmann}, {Tholen}, {Meech}, \&
  {Cruikshank}}]{Hart+90}
{Hartmann}, W.~K., {Tholen}, D.~J., {Meech}, K.~J., \& {Cruikshank}, D.~P.
  1990, Icarus, 83, 1

\bibitem[{{Jewitt} {et~al.}(2007){Jewitt}, {Peixinho}, \&
  {Hsieh}}]{JewPeiHsi07}
{Jewitt}, D., {Peixinho}, N., \& {Hsieh}, H.~H. 2007, Astron. J., 134, 2046

\bibitem[{{Jewitt} \& {Luu}(2001)}]{JewLuu01}
{Jewitt}, D.~C., \& {Luu}, J.~X. 2001, Astron. J., 122, 2099

\bibitem[{{Kolmogorov}(1933)}]{Kolmogorov33}
{Kolmogorov}, A.~N. 1933, Giornale dell' Istituto Italiano degli Attuari, 4, 83

\bibitem[{{Kuchner} {et~al.}(2002){Kuchner}, {Brown}, \& {Holman}}]{Kuc+02}
{Kuchner}, M., {Brown}, M., \& {Holman}, M. 2002, Astron. J., 124, 1221

\bibitem[{{Levenberg}(1944)}]{Lev44}
{Levenberg}, K. 1944, Quart. Appl. Math.

\bibitem[{{Levison} \& {Stern}(2001)}]{LS01}
{Levison}, H.~F., \& {Stern}, S.~A. 2001, Astrophys. J., 121, 1730

\bibitem[{Luu \& Jewitt(1996)}]{LuuJew96}
Luu, J., \& Jewitt, D. 1996, Astron. J., 112, 2310

\bibitem[{{Lykawka} \& {Mukai}(2008)}]{LykMuk08}
Lykawka, P.~S., \& Mukai, T.\ 2008, Astron. J., 135, 116.

\bibitem[{{Lykawka} \& {Mukai}(2007)}]{LykMuk07}
---. 2007, Icarus, 189, 213

\bibitem[{{Malhotra}(1995)}]{Malho95}
{Malhotra}, R. 1995, Astrophys. J., 110, 420

\bibitem[{{Marquardt}(1963)}]{Mar63}
{Marquardt}, D. 1963, SIAM J. Appl. Math., 11, 431.

\bibitem[{{McBride} {et~al.}(2003){McBride}, {Green}, {Davies}, {Tholen},
  {Sheppard}, {Whiteley}, \& {Hillier}}]{McB+03}
{McBride}, N., {Green}, S.~F., {Davies}, J.~K., {Tholen}, D.~J., {Sheppard},
  S.~S., {Whiteley}, R.~J., \& {Hillier}, J.~K. 2003, Icarus, 161, 501

\bibitem[Morbidelli et~al.(2003)]{Morb+03} Morbidelli, A., Brown, M.~E., \& Levison, H.~F.\ 2003, Earth Moon and Planets, 92, 1 

\bibitem[{{Noll} {et~al.}(2008){Noll}, {Grundy}, {Stephens}, {Levison}, \&
  {Kern}}]{Noll+08}
{Noll}, K.~S., {Grundy}, W.~M., {Stephens}, D.~C., {Levison}, H.~F., \& {Kern},
  S.~D. 2008, Icarus

\bibitem[{{Peixinho} {et~al.}(2004){Peixinho}, {Boehnhardt}, {Belskaya},
  {Doressoundiram}, {Barucci}, \& {Delsanti}}]{Peix+04}
{Peixinho}, N., {Boehnhardt}, H., {Belskaya}, I., {Doressoundiram}, A.,
  {Barucci}, M., \& {Delsanti}, A. 2004, Icarus, 170, 153

\bibitem[{{Romanishin} \& {Tegler}(1999)}]{RomTeg99}
{Romanishin}, W., \& {Tegler}, S.~C. 1999, Nature, 398, 129

\bibitem[{{Smirnov}(1939)}]{Smirnov39}
{Smirnov}, N.~V. 1939, Bull. Moscow Univ., 2, 3

\bibitem[{{Spearman}(1904)}]{Spe04}
{Spearman}, C. 1904, Am. J. Psychol., 57, 72

\bibitem[{Tegler \& Romanishin(2000)}]{TegRom00}
Tegler, S., \& Romanishin, W. 2000, Nature, 407, 979

\bibitem[{{Tegler} \& {Romanishin}(1998)}]{TegRom98}
--- . 1998, Nature, 392, 49

\bibitem[{{Tegler} {et~al.}(2003){Tegler}, {Romanishin}, \&
  {Consolmagno}}]{TRC03}
{Tegler}, S.~C., {Romanishin}, W., \& {Consolmagno}, S.~J. 2003,
  Astrophys. J., Lett., 599, L49

\bibitem[{{Th{\' e}bault} \& {Doressoundiram}(2003)}]{TheDor03}
{Th{\' e}bault}, P., \& {Doressoundiram}, A. 2003, Icarus, 162, 27

\bibitem[{{Trujillo} \& {Brown}(2002)}]{TruBro02}
{Trujillo}, C.~A., \& {Brown}, M.~E. 2002, Astrophys. J., Lett., 566, L125

\bibitem[{{Wilcoxon}(1945)}]{Wilcoxon45}
{Wilcoxon}, F. 1945, Biometrika, 1, 80

\end{thebibliography}
\end{document}